\documentclass[11pt]{article}
\author{G.~Gubbiotti$^1$ \and M.~C.~Nucci$^2$}
\title{Are all classical superintegrable systems in  two-dimensional space linearizable?}
\date{$^1$Dipartimento di Matematica e Fisica, Universit\`{a} degli Studi Roma Tre,
 \& INFN Sezione di Roma Tre, 00146 Roma, Italy\\ [0.2cm]
$^2$Dipartimento di Matematica e Informatica, Universit\`a degli Studi di
Perugia, \& INFN Sezione di Perugia, 06123 Perugia, Italy}

\usepackage{amsmath}
\usepackage{amssymb}
\usepackage{mathtools}
\usepackage{braket}
\usepackage{cite}

\newcommand{\half}{\frac{1}{2}}
\newcommand{\I}{\mathrm{I}}
\newcommand{\II}{\mathrm{II}}
\newcommand{\III}{\mathrm{III}}
\newcommand{\IV}{\mathrm{IV}}

\newcommand{\ud}{\mathrm{d}}
\newcommand{\der}[3][]{\frac{\ud^{#1} #2}{\ud #3^{#1}}}

\newcommand{\Sl}{\mathrm{sl}}
\newcommand{\R}{\mathbb{R}}
\newcommand{\di}{\displaystyle}
\DeclarePairedDelimiter{\Span}{\langle}{\rangle} \allowdisplaybreaks

\newcounter{rmk}
\renewcommand{\thermk}{\arabic{rmk}}
\newenvironment{remark}%
{\refstepcounter{rmk}\noindent \textbf{Remark
\thermk:}}{\hfill$\blacksquare$\\\par \noindent}

\begin{document}

\maketitle

\begin{abstract}
    Several examples of classical superintegrable systems in two-dimensional space
    are shown to possess hidden symmetries leading to their linearization. They are those determined 50 years ago in
    \cite{Fris1965}, and the more recent Tremblay-Turbiner-Winternitz system \cite{TTW}.
     We conjecture that all classical superintegrable systems in two-dimensional space
     have hidden symmetries that make them linearizable.
\end{abstract}

\section{Introduction}
About 50 years ago in a seminal paper \cite{Fris1965}, the authors  considered
Hamiltonian systems with Hamiltonian given either in cartesian coordinates,
i.e.:
\begin{equation}
    H = \half \left( p_{1}^{2} + p_{2}^{2} \right) + V(x_{1},x_{2}),
        \label{eqn:canhamc}
\end{equation}
or in polar coordinates, i.e.:
\begin{equation}
    H = \half \left( p_{r}^{2} + \frac{p_{\varphi}^{2}}{r^2} \right) + V(r,\varphi),
        \label{eqn:canhamp}
\end{equation}
 Their purpose was to determine all the
potentials such that the corresponding Hamiltonian system admits two first
integrals that are quadratic in the momenta,  in addition to the Hamiltonian.
 No assumption about the separation of variables in the Hamilton-Jacobi  equation
was made a priori. Four independent potentials were found and it was proven
that the corresponding Hamilton-Jacobi  equation was separable in at least two
different coordinate systems. Two of the four potentials were given in
cartesian coordinates:
\begin{subequations}
    \begin{align}
        V_{\I}(x_{1},x_{2}) &= \frac{\omega^2}{2} (x_{1}^{2} + x_{2}^{2})
        + \frac{\beta_{1}}{x_{1}^{2}} + \frac{\beta_{2}}{x_{2}^{2}},
        \label{eqn:vi}
        \\
        V_{\II}(x_{1},x_{2}) &= \frac{\omega^2}{2} (4 x_{1}^{2} + x_{2}^{2}) + \beta_{1}x_{1}
        + \frac{\beta_{2}}{x_{2}^{2}},
        \label{eqn:vii}
      \end{align}
while the other two were given in polar coordinates:
    \begin{align}
        V_{\III}(r,\varphi) &= \frac{\alpha}{r} + \frac{1}{r^{2}}
        \left( \frac{\beta_{1}}{\cos^{2}\left( \frac{\varphi}{2} \right)}+
        \frac{\beta_{2}}{\sin^{2}\left( \frac{\varphi}{2} \right)},
        \right)
        \label{eqn:viii}
        \\
        V_{\IV}(r,\varphi) &=
        \frac{\alpha}{r} + \frac{1}{\sqrt{r}}
        \left( \beta_{1} \cos\left( \frac{\varphi}{2}\right)
        +\beta_{2} \sin\left( \frac{\varphi}{2} \right)\right).
        \label{eqn:viv}
    \end{align}
    \label{eqn:vs}
\end{subequations}
 These four cases belong to the class of two-dimensional
superintegrable systems, namely those Hamiltonian systems that admit three
first integrals. Actually their are also maximally superintegrable. In fact a
Hamiltonian system with $n$ degrees of freedom is called superintegrable if
allows $n+1$ integrals, and maximally superintegrable if the integrals are $2n
- 1$. For $n = 2$ the two definitions coincide.

In any undergraduate text of Mechanics, e.g. \cite{Goldstein}, it is shown that
Kepler problem in polar coordinates is linearizable, namely that one can
exactly transform its nonlinear equations of motion into the equation of an
harmonic linear oscillator. In \cite{harmony} it was shown that such a
linearization can be achieved by means of the reduction method that was
proposed in \cite{kepler} in order to find hidden symmetries of Kepler problem. 
Moreover,  the reduction method was successfully
applied to generalizations of the Kepler problem with and without drag in order
to find their hidden linearity, although not all of them admit a Lagrangian
description \cite{GorLeach87}.

In 2009 a new
two-dimensional superintegrable system was determined \cite{TTW}, and  it has
been known since as the Tremblay-Turbiner-Winternitz (TTW) system.

In 2011  a two-dimensional superintegrable system  such that the corresponding
Hamilton-Jacobi  equation does not admit separation of variables in any
coordinates was studied in \cite{PostWint11}. In \cite{NucPost} it was found
that its Lagrangian equations can be transformed into a linear third-order
equation by applying the reduction method \cite{kepler}.

In the present paper we show that the Lagrangian equations corresponding to the
potentials $V_{\I}, V_{\II}, V_{\III}$, and some of their generalizations are
all linearizable by means of their hidden symmetries.  We also prove that the TTW system is linearizable.

The Hamiltonian system with potential  $V_{\IV}$ is a subcase of the
linearizable cases determined in \cite{harmony}, where the following Newtonian
equations were considered:
\begin{eqnarray}
\ddot r-r\dot {\varphi}^2+g=0,\\
r\ddot \varphi +2\dot r\dot {\varphi}+h=0,
\end{eqnarray}
with
\begin{equation}
g=\frac{U''(\varphi)+U(\varphi)}{r^2}+2\frac{W'(\varphi)}{r^{3/2}},\quad\quad
h=\frac{W(\varphi)}{r^{3/2}}.
\end{equation}
 It corresponds to the substitution \cite{GorLeach87}:
\begin{equation} U=\alpha,\quad \quad W=\frac{1}{2}\left( \beta_1\sin\left(
\frac{\varphi}{2} \right)- \beta_2\cos\left( \frac{\varphi}{2}
\right)\right).\end{equation} See also \cite{Sen87}.

 All the superintegrable systems that we consider here are in real
Euclidean space. In a forthcoming paper \cite{2Dnonflat}, we will show that
 many known superintegrable systems in space of non-constant curvature  are also
linearizable, e.g.  the three superintegrable systems for
the Darboux space of Type I determined in \cite{KalKreWint}.

We conclude with a conjecture, namely that all two-dimensional superintegrable
systems are linearizable by means of their hidden symmetries.

\section{Linearity of the Lagrangian equations with potentials $V_{\I}$,
 $V_{\II}$, and $V_{\III}$}
The Lagrangian corresponding to the Hamiltonian \eqref{eqn:canhamc} in
cartesian coordinates is:
\begin{equation}
        L = \half \left( \dot{x}_{1}^{2} + \dot{x}_{2}^{2} \right) -
        V(x_{1},x_{2}),
    \label{eqn:canlagr}
\end{equation}
while the Lagrangian corresponding to the Hamiltonian \eqref{eqn:canhamp} in
polar coordinates  is
\begin{equation}
L=\half(\dot r^2+r^2\dot\varphi^2)-V(r,\varphi).\label{Lagrpol}
\end{equation}
\begin{remark} We have applied Douglas' method \cite{Douglas} to the
Lagrangian equations corresponding to the four potentials. The two potentials
$V_{\I}$ and $V_{\II}$ lead to many different Lagrangians, while in the case of
potentials $V_{\III}$ and $V_{\IV}$, there exists only one Lagrangian in
analogy with Kepler's problem. \end{remark}

\subsection{The potential $V_I$}

The Lagrangian equations corresponding to the Lagrangian \eqref{eqn:canlagr}
with  $V=V_{\I}$  are:
\begin{eqnarray}
        \ddot{x}_{1}= - \omega^2 x_{1} + \frac{2\beta_{1}}{x_{1}^{3}},
       \nonumber
        \\
        \ddot{x}_{2}= - \omega^2 x_{2} + \frac{2\beta_{2}}{x_{2}^{3}}.
            \label{eqn:vieq}
\end{eqnarray}
This Lagrangian admits three Noether symmetries generated by the operators:
\begin{equation}
    \begin{gathered}
        \Sigma_{1}=\partial_{t}, \quad
        \Sigma_{2}=\cos(2\omega t)\partial_{t}
        - \omega x_1\sin(2\omega t)\partial_{x_1}-\omega x_2\sin(2\omega t)\partial_{x_2},
        \\
        \Sigma_{3}=\sin(2\omega t)\partial_{t}
        + \omega x_1\cos(2\omega t)\partial_{x_1}+\omega x_2\cos(2\omega
        t)\partial_{x_2}.
    \end{gathered}
    \label{noethersymV1}
\end{equation}
which correspond to the algebra $sl(2,\mathbb{R})$. The application of
Noether's theorem yields three first integrals. From $\Sigma_1$ comes the
Hamiltonian, i.e.:
\begin{equation}
    H_I = \half \left(\dot{x}_{1}^{2} + \dot{x}_{2}^{2} \right)
    + \frac{\omega^2}{2} (x_{1}^{2} + x_{2}^{2})
        + \frac{\beta_{1}}{x_{1}^{2}} + \frac{\beta_{2}}{x_{2}^{2}}.
    \label{hamI}
\end{equation}
and from $\Sigma_2$ and $\Sigma_3$ the following two time-dependent integrals:
\begin{eqnarray}
        K_{2} &=&
        \left[\frac{\beta_{1}}{x_{1}^2} + \frac{\beta_{2}}{x_{2}^2}
        + \half(\dot{x}_1^2 + \dot{x}_2^2)
        -\frac{\omega^2}{2} (x_{1}^2 + x_{2}^2) \right] \cos(2 \omega t)
    \nonumber    \\
        &&+\omega(x_{1} \dot{x}_1 + x_{2} \dot{x}_2) \sin(2 \omega t), \label{fintvi2}
         \end{eqnarray}
and
        \begin{eqnarray}
        K_{3} &=&
        \left[\frac{\beta_{1}}{x_{1}^2} + \frac{\beta_{2}}{x_{2}^2}
        + \half(\dot{x}_1^2 + \dot{x}_2^2)
        -\frac{\omega^2}{2} (x_{1}^2 + x_{2}^2) \right] \sin(2 \omega t) \nonumber  \\
        &&-\omega(x_{1} \dot{x}_1 + x_{2} \dot{x}_2) \cos(2 \omega t),
               \label{fintvi3}
\end{eqnarray}
respectively.\\

\begin{remark} Another time-independent first integral can
be obtained by the following ubiquitous combination:
\begin{equation}
 H_I^2-K_{2}^{2} - K_{3}^{2} =\omega^2\left( 2\beta_1+2\beta_2
 +2\beta_1\frac{x_2^2}{x_1^2}
  +2\beta_2\frac{x_1^2}{x_2^2}+(x_2\dot x_1-x_1\dot x_2)^2\right)
    \label{eqn:fintnotime}
\end{equation}
Such a combination can be found in other instances where a couple of
time-dependent first integrals are derived from Noether's theorem.
\end{remark}
The presence of the algebra $sl(2,\mathbb{R})$ suggests to eliminate the two
parameters $\beta_1$ and $\beta_2$ by raising the order, as it was done in
\cite{Leach2003} in the case of the isotonic oscillator.   We solve system
\eqref{eqn:vieq} with respect to $\beta_1$ and $\beta_2$, i.e.:
\begin{eqnarray}
\beta_1=\half(x_1^3 \ddot x_1+ \omega^2 x_1^4), \nonumber \\
\beta_2=\half(x_2^3 \ddot x_2+ \omega^2 x_2^4),
\end{eqnarray}
and then we differentiate them with respect to $t$  in order to get the
following two third-order equations:
\begin{eqnarray}
            \dddot{x}_{1}&=-\displaystyle\frac{ \dot{x}_{1}}{x_1} ( 4 \omega^2 x_1 + 3 \ddot{x}_{1} ) ,
        \nonumber
        \\
           \dddot{x}_{2}&=-\displaystyle\frac{ \dot{x}_{2}}{x_2} ( 4 \omega^2 x_2 + 3 \ddot{x}_{2} ).
              \label{eqn:vieq3ord}
\end{eqnarray}
This system admits a thirteen-dimensional Lie point symmetry algebra generated
by the following operators:
\begin{eqnarray}
\Gamma_{1} &=& \cos(2 \omega t)\partial_t- \sin(2 \omega t) \omega
x_1\partial_{x_1}- \sin(2 \omega t) \omega x_2\partial_{x_2},
   \nonumber     \\
\Gamma_{2} &=&\sin(2 \omega t)\partial_t + \cos(2 \omega t) \omega
x_1\partial_{x_1}+\cos(2 \omega t) \omega x_2\partial_{x_2} ,
     \nonumber     \\
        \Gamma_{3} &=& \partial_{t},
    \nonumber      \\
        \Gamma_{4} &=&\frac{\cos(2 \omega t)}{x_1}\,\partial_{x_1},
\nonumber
        \\
        \Gamma_{5} &=&\frac{\sin(2 \omega t)}{x_1}\,\partial_{x_1},
\nonumber
        \\
        \Gamma_{6} &=&\frac{\cos(2 \omega t)}{x_2}\,\partial_{x_2},
 \nonumber         \\
        \Gamma_{7} &=&\frac{\sin(2 \omega t)}{x_2}\,\partial_{x_2},
 \label{eqn:symmvieqd}       \\
        \Gamma_{8} &=&\frac{x_2^2}{x_1}\, \partial_{x_1},
\nonumber          \\
        \Gamma_{9} &=&x_1\,\partial_{x_1},
 \nonumber         \\
        \Gamma_{10} &=&\frac{1}{x_1}\, \partial_{x_1},
  \nonumber        \\
        \Gamma_{11} &=&\frac{x_1^2}{x_2} \,\partial_{x_2},
 \nonumber         \\
        \Gamma_{12} &=&x_2\partial_{x_2},
 \nonumber         \\
        \Gamma_{13} &=&\frac{1}{x_2}\, \partial_{x_2}.
 \nonumber
\end{eqnarray}
Therefore system \eqref{eqn:vieq3ord} is linearizable.
%
In order to find the linearizing transformation we could use the method in
\cite{Soh2001,Popovich2003} based on the classification of the four-dimensional
Abelian subalgebras \cite{PateraWint}. Instead we recall  that the following
linear system\footnote{It is the derivative with respect to $t$ of the
equations of a two-dimensional isotropic oscillator with frequency $2\omega$.}:
\begin{eqnarray}
\dddot u_1&=&-4\omega^2\dot u_1,\nonumber \\
\dddot u_2&=&-4\omega^2\dot u_2,\label{dh}
\end{eqnarray}
admits a thirteen-dimensional Lie point symmetry algebra generated by the
following operators:
\begin{eqnarray}
\overline\Gamma_{1} &=& \cos(2 \omega t)\partial_t- 2\sin(2 \omega t) \omega
u_1\partial_{u_1}- 2\sin(2 \omega t) \omega u_2\partial_{u_2},
   \nonumber     \\
\overline\Gamma_{2} &=&\sin(2 \omega t)\partial_t + 2\cos(2 \omega t) \omega
u_1\partial_{u_1}+2\cos(2 \omega t) \omega u_2\partial_{u_2} ,
     \nonumber     \\
 \overline       \Gamma_{3} &=& \partial_{t},
    \nonumber      \\
   \overline     \Gamma_{4} &=&\cos(2 \omega t)\,\partial_{u_1},
\nonumber
        \\
  \overline      \Gamma_{5} &=&\sin(2 \omega t)\,\partial_{u_1},
\nonumber
        \\
   \overline     \Gamma_{6} &=&\cos(2 \omega t)\,\partial_{u_2},
 \nonumber         \\
     \overline   \Gamma_{7} &=&\sin(2 \omega t)\,\partial_{u_2},
 \label{eqn:symmdh}       \\
  \overline      \Gamma_{8} &=&u_2\, \partial_{u_1},
\nonumber          \\
  \overline      \Gamma_{9} &=&{u_1}\, \partial_{u_1},
 \nonumber         \\
   \overline     \Gamma_{10} &=&\partial_{u_1},
  \nonumber        \\
    \overline    \Gamma_{11} &=&u_1\,\partial_{u_2},
 \nonumber         \\
    \overline    \Gamma_{12} &=&{u_2}\, \partial_{u_2},
 \nonumber         \\
   \overline     \Gamma_{13} &=&\partial_{u_2}.
 \nonumber
\end{eqnarray}
Consequently, if we make the following transformation:
\begin{equation}
    u_{1} = \di\frac{x_{1}^{2}}{2}, \quad\quad u_{2}= \di\frac{x_{2}^{2}}{2}
    \label{eqn:changevi2}
\end{equation}
system \eqref{eqn:vieq}  becomes the linear system \eqref{dh}.\\

More recently the following generalization of the potential $V_{\I}$
 has been proposed and proved superintegrable
\cite{Rodriguez2008,Rodriguez2009,Ranada2010,Ranada2004}:
\begin{equation}
    V_{\I}^{{\rm gen}}(x_{1},x_{2}) = \frac{\omega_{1}^2}{2} x_{1}^{2} +
    \frac{\omega_{2}^2}{2} x_{2}^{2}
    + \frac{\beta_{1}}{x_{1}^{2}} + \frac{\beta_{2}}{x_{2}^{2}}.
    \label{eqn:vial1al2}
\end{equation}
Applying the same procedure as described above to the corresponding Lagrangian
equations, i.e.:
\begin{eqnarray}
        \ddot{x}_{1}&=&-\omega_1^2 x_{1} + \frac{2\beta_{1}}{x_{1}^{3}},
       \nonumber
        \\
        \ddot{x}_{2}&=& -\omega_{2}^2 x_{2} + \frac{2\beta_{2}}{x_{2}^{3}},
            \label{eqn:vieqal1al2}
\end{eqnarray}
yields the following system of two third-order equations:
\begin{eqnarray}
 \dddot{x}_{1}&=&- \frac{\dot{x}_{1}}{x_1} ( 4 \omega_{1}^2 x_1 + 3 \ddot{x}_{1} ),
        \nonumber
        \\
\dddot{x}_{2}&=&- \frac{\dot{x}_{2}}{x_{2}} ( 4 \omega_{2}^2 x_2 + 3
\ddot{x}_{2} ),
          \label{eqn:vieqdal1al2}
\end{eqnarray}
which admits a  nine-dimensional Lie point symmetry algebra generated by the
 operators $\Gamma_3,\Gamma_4,\Gamma_5,\Gamma_6,\Gamma_7,\Gamma_8,\Gamma_{10},
 \Gamma_{11}, \Gamma_{13}$ in \eqref{eqn:symmvieqd}.
Indeed by applying again the transformation \eqref{eqn:changevi2} we obtain
that the system \eqref{eqn:vieqdal1al2} is transformed into the following
linear system:
\begin{eqnarray}
    \dddot{u}_{1} &=& -4\omega_1^2 \dot{u}_{1}, \nonumber\\
    \dddot{u}_{2} &=& -4\omega_2^2 \dot{u}_{2},
    \label{eqn:vial1al2linearization2}
\end{eqnarray}
namely the derivative with respect to $t$ of the equations of a two-dimensional
anisotropic oscillator.

\subsection{The potential $V_{\II}$}

The Lagrangian equations corresponding to the Lagrangian \eqref{eqn:canlagr}
with  $V=V_{\II}$  are:
\begin{subequations}
    \begin{align}
        \ddot{x}_{1}&=- 4 \omega^{2} x_{1} -\beta_{1},
        \label{eqn:viieqa}
        \\
        \ddot{x}_{2}&=-  \omega^{2} x_{2} + \frac{2\beta_{2}}{x_{2}^{3}}.
        \label{eqn:viieqb}
    \end{align}
    \label{eqn:viieq}
\end{subequations}
This Lagrangian admits three Noether symmetries generated by the following
operators:
\begin{equation}
    \Upsilon_{1} = \partial_{t}, \quad
    \Upsilon_{2} = \sin(2\omega t)\partial_{x_{1}}, \quad
    \Upsilon_{3} = \cos(2\omega t)\partial_{x_{2}},
    \label{eqn:symmnothvii}
\end{equation}
that is the algebra $A_{3,6} \simeq
\Span{\Upsilon_{1}/(2\omega),\Upsilon_{3},\Upsilon_{2}}$ in the classification
given in \cite{PateraWint}. The application of Noether's theorem yields three
first integrals. From $\Upsilon_{1}$ comes the Hamiltonian, i.e.:
\begin{equation}
H_{\II}=\half \left( \dot{x}_{1}^{2} + \dot{x}_{2}^{2}
\right)+\frac{\omega^2}{2} (4 x_{1}^{2} + x_{2}^{2}) + \beta_{1}x_{1}
        + \frac{\beta_{2}}{x_{2}^{2}} \label{HamII}
\end{equation}
and from $\Upsilon_{2}$ and $\Upsilon_3$ the following two time-dependent
integrals:
\begin{equation}
        Y_{2} =\cos(2\omega t) \beta_{1} + 4\cos(2\omega t) \omega^2 x_1
        - 2\sin(2 \omega t) \omega \dot{x}_{1},
        \label{eqn:fivii2}
\end{equation}
and \begin{equation}
 Y_{3} =- 2\cos(2\omega t)\omega \dot{x}_{1} - \sin(2 \omega t) \beta_{1}
        - 4 \sin(2 \omega t) \omega^2 x_{1},
        \label{eqn:fivii3}
       \end{equation}
       respectively.\\

\begin{remark} The following combination of \eqref{eqn:fivii2} and \eqref{eqn:fivii3}
yields the Hamiltonian for the equation \eqref{eqn:viieqa} only, i.e.:
\begin{equation}
  H_{1}=\frac{Y_{2}^{2} + Y_{3}^{2}}{8\omega^{2}} =
    \half\dot{x}_{1}^{2} + 2\omega^{2} x_{1}^{2} + \beta_{1} x_{1}
    + \frac{\beta_{1}^{2}}{8\omega^{2}}.
    \label{eqn:y23toh1}
\end{equation}
The following combination of \eqref{eqn:y23toh1} and the Hamiltonian
\eqref{HamII} yields the Hamiltonian for the equation \eqref{eqn:viieqb} only,
i.e.:
\begin{equation}
   H_{2}= H_{\II} - \frac{Y_{2}^{2} + Y_{3}^{2}}{8\omega^{2}} =
    \half\dot{x}_{2}^{2} + \half\omega^{2} x_{2}^{2}
    + \frac{\beta_{2}}{x_{2}^{2}} - \frac{\beta_{1}^{2}}{8\omega^{2}}.
    \label{eqn:y23toh2}
\end{equation}
Of course, the addition/sottraction of the constant
$\di\frac{\beta_{1}^{2}}{8\omega^{2}}$ does not influence either the
Hamiltonian $H_1$ or $H_2$.
\end{remark}
We solve system \eqref{eqn:viieqa}-\eqref{eqn:viieqb} with respect to $\beta_1$
and $\beta_2$, i.e.:
\begin{eqnarray}
\beta_1&=&- \ddot x_1-4 \omega^2 x_1, \nonumber \\
\beta_2&=&\half( x_2^3\ddot x_2+ \omega^2 x_2^4),
\end{eqnarray}
and then we take the derivative with respect to $t$  in order to get the
following system of two third-order equations:
\begin{eqnarray}
\dddot{x}_{1}&=&-4 \omega^2 \dot{x}_{1},\nonumber   \\
\dddot{x}_{2}&=&-\displaystyle\frac{ \dot{x}_{2}}{x_2} ( 4 \omega^2 x_2 + 3
\ddot{x}_{2} ).              \label{eqn:viieq3ord}
\end{eqnarray}
It should not be a surprise that this system admits a thirteen-dimensional Lie
symmetry algebra.
Consequently, the transformation \begin{equation} u_1=x_1,\quad \quad
u_2=\frac{x_2^2}{2}\end{equation} takes system \eqref{eqn:viieq3ord} into the
linear system \eqref{dh}, namely that obtained by taking the derivative with
respect to $t$ of the equations of a two-dimensional isotropic oscillator with
frequency $2\omega$.

\subsection{The potential $V_{\III}$}
The Lagrangian equations corresponding to the Lagrangian \eqref{Lagrpol} with
 $V=V_{\III}$  are:
\begin{eqnarray}
     \ddot{r}&=&
        \frac{\alpha}{r^{2}} +r\dot{\varphi}^2 + \frac{2}{r^{3}}
        \left( \frac{\beta_{1}}{\cos^{2}\left( \frac{\varphi}{2} \right)}+
        \frac{\beta_{2}}{\sin^{2}\left( \frac{\varphi}{2} \right)}
        \right),
        \nonumber
        \\
        \ddot{\varphi} &=& -  \frac{2}{r}\dot{r}\dot{\varphi}-
        \frac{1}{r^{4}}
        \left( \frac{\beta_{1}\sin\left(\frac{\varphi}{2} \right)}{%
            \cos^{3}\left( \frac{\varphi}{2} \right)}-
        \frac{\beta_{2}\cos\left( \frac{\varphi}{2}\right)}{%
            \sin^{3}\left( \frac{\varphi}{2} \right)}
        \right).
                \label{eqn:viiieq}
\end{eqnarray}
This Lagrangian admits one Noether symmetry, i.e. translation in $t$, and
Noether theorem yields the Hamiltonian.\\
We now write the two second-order Lagrangian equations  \eqref{eqn:viiieq} as
the following four first-order equations
\begin{eqnarray}
        \dot{w}_{1} &=& w_{3},
       \nonumber
        \\
        \dot{w}_{2} &=& w_{4},
       \nonumber        \\
        \dot{w}_{3}&=&\frac{\alpha}{w_1^{2}} +w_1w_4^2 + \frac{2}{w_1^{3}}
        \left( \frac{\beta_{1}}{\cos^{2}\left( \frac{w_2}{2} \right)}+
        \frac{\beta_{2}}{\sin^{2}\left( \frac{w_2}{2} \right)}
        \right),
           \label{eqn:viii1ord}     \\
        \dot{w_4} &=& -  \frac{2}{w_1}w_3w_4-
        \frac{1}{w_1^{4}}
        \left( \frac{\beta_{1}\sin\left(\frac{w_2}{2} \right)}{%
            \cos^{3}\left( \frac{w_2}{2} \right)}-
        \frac{\beta_{2}\cos\left( \frac{w_2}{2}\right)}{%
            \sin^{3}\left( \frac{w_2}{2} \right)}
        \right),
      \nonumber
    \end{eqnarray}
with the identification \begin{equation}
    (w_{1},w_{2},w_{3},w_{4})\equiv(r,\varphi,\dot{r},\dot{\varphi}).
    \label{eqn:syssubs}
\end{equation}
We apply the reduction method \cite{kepler} by choosing $w_{2}=y$ as the new
independent variable, and consequently the following system of three
first-order equations is obtained:
\begin{eqnarray}
        \der{w_{1}}{y} &=& \frac{w_{3}}{w_4},
        \label{eqn:viii3eq_1}
         \\
        \der{w_{3}}{y}&=&\frac{\alpha}{w_1^{2}w_4} +w_1w_4 + \frac{2}{w_1^{3}w_4}
        \left( \frac{\beta_{1}}{\cos^{2}\left( \frac{y}{2} \right)}+
        \frac{\beta_{2}}{\sin^{2}\left( \frac{y}{2} \right)}
        \right),
           \label{eqn:viii3eq_2}     \\
       \der{w_{4}}{y} &=& -  \frac{2}{w_1}w_3-
        \frac{1}{w_1^{4}w_4}
        \left( \frac{\beta_{1}\sin\left(\frac{y}{2} \right)}{%
            \cos^{3}\left( \frac{y}{2} \right)}-
        \frac{\beta_{2}\cos\left( \frac{y}{2}\right)}{%
            \sin^{3}\left( \frac{y}{2} \right)}
        \right),
       \label{eqn:viii3eq_3}
    \end{eqnarray}
 We derive $w_3$ from equation  \eqref{eqn:viii3eq_1}, i.e.,
\begin{equation}
    w_{3} = w_{4} \der{w_{1}}{y},
    \label{eqn:w3sol}
\end{equation}
and consequently equation \eqref{eqn:viii3eq_3} becomes:
\begin{equation}
   \der{w_{4}}{y} +  \frac{2w_{4}}{w_1}\der{w_{1}}{y} +
    \frac{1}{w_1^4w_4}\left( \frac{\beta_{1}\sin\left( \frac{y}{2} \right)}{
        \cos^{3}\left(\frac{y}{2} \right)}-
        \frac{\beta_{2}\cos\left( \frac{y}{2} \right)}{%
            \sin^{3}\left( \frac{y}{2} \right)}
    \right)=0,
    \label{eqn:firstint}
\end{equation}
which can be simplified by means of the following transformation, i.e.:
\begin{equation}
w_4=\frac{r_4}{w_1^2},
    \label{eqn:u2def}
\end{equation}
with $r_4$ a new function of $y$ that then has to satisfy the following
equation:
\begin{equation}
 \der{r_{4}}{y} =-\frac{1}{r_4}\left(\frac{\beta_{1}\sin\left( \frac{y}{2} \right)}{
        \cos^{3}\left(\frac{y}{2} \right)}-\frac{\beta_{2}
        \cos\left( \frac{y}{2} \right)}{\sin^{3}\left( \frac{y}{2} \right)}
    \right).
    \end{equation}
Its general solution is easily obtained to be:
\begin{equation}
r_4=\pm\sqrt{a_1 - 2\left(\frac{\beta_{1}\sin\left( \frac{y}{2} \right)}{
        \cos\left(\frac{y}{2} \right)}+\frac{\beta_{2}
        \cos\left( \frac{y}{2} \right)}{\sin\left( \frac{y}{2}
        \right)}\right)},
\end{equation}
with  $a_1$ an arbitrary constant. Finally, equation \eqref{eqn:viii3eq_2}
becomes the following second-order differential equation:
\begin{eqnarray}
    \der[2]{w_{1}}{y} &\!\!\!=&\!\!\! \frac{2}{w_{1}}\left( \der{w_{1}}{y} \right)^{2}
 +
\frac{w_1(\alpha w_1 + a_1)\sin^2\left( \frac{y}{2} \right)\cos^2\left(
\frac{y}{2} \right)}{a_1\sin^2\left( \frac{y}{2} \right)\cos^2\left(
\frac{y}{2} \right)- 2\beta_{1}\sin^2\left( \frac{y}{2} \right)- 2\beta_{2}
        \cos^2\left( \frac{y}{2} \right)}   \nonumber \\\!\!\!&&\!\!\!\!\!\!+
   \frac{\beta_{1}\sin^4\left( \frac{y}{2} \right) -\beta_{2}
        \cos^4\left( \frac{y}{2} \right)}
{\sin\left( \frac{y}{2} \right)\cos\left(\frac{y}{2}\right)
\left(a_1\sin^2\left( \frac{y}{2} \right)\cos^2\left( \frac{y}{2} \right) -
2\beta_{1}\sin^2\left( \frac{y}{2} \right)- 2\beta_{2}
        \cos^2\left( \frac{y}{2} \right) \right)}   \der{w_{1}}{y} .
    \label{eqn:w1fin}
\end{eqnarray}
This equation admits an eight-dimensional Lie point symmetry algebra, which
means that it is linearizable. The linearizing transformation is obtained by means
of Lie's canonical representation of a two-dimensional abelian intransitive
subalgebra \cite{Lie12}. One such subalgebra is that generated by  the
following two operators:
\begin{eqnarray}
\Xi_1 &=& (2\beta_1-2\beta_2-\cos(y)a_1)w_1^2\partial_{w_1},\nonumber\\
\Xi_2&=&\sqrt{4\beta_2\cos(y)-4\beta_1\cos(y)+a_1\cos^2(y)+4\beta_1+4\beta_2-a_1}w_1^2\partial_{w_1},
\end{eqnarray}
that we have to put in the canonical form  $\partial_{\tilde w_1}, \tilde y\partial_{\tilde w_1}$. Therefore the transformation
\begin{eqnarray}
\tilde y&=& -\frac{\sqrt{4\beta_2\cos(y)-4\beta_1\cos(y)+a_1\cos^2(y)+4\beta_1+4\beta_2-a_1}}{-2\beta_1+2\beta_2+\cos(y)a_1},\\
\tilde w_1&=&\frac{1}{(-2\beta_1+2\beta_2+\cos(y)a_1)w_1}\label{w1tr}
\end{eqnarray}
takes equation \eqref{eqn:w1fin} into a linear equation of the type
\begin{equation}
\frac{{\rm d}^2\tilde w_1}{{\rm d} \tilde y^2}=\mathfrak{F}(\tilde y).
\end{equation}
Actually \eqref{w1tr} suggests the simpler transformation \begin{equation} u=\frac{1}{w_1},\label{1su}\end{equation} that applied to equation \eqref{eqn:w1fin} yields  the following linear  equation:
\begin{equation}
\ddot  u=\frac{ \dot u\left(\beta_1\sin^4\left(
\frac{y}{2} \right)-\beta_2\cos^4\left(
\frac{y}{2} \right)\right)-(\alpha-a_1u)\sin^3\left(
\frac{y}{2} \right)\cos^3\left(
\frac{y}{2} \right)}{\left(a_1\sin^2\left(\frac{y}{2} \right)\cos^2\left(\frac{y}{2} \right)-2\beta_1\sin^2\left(\frac{y}{2} \right)-2\beta_2\cos^2\left(\frac{y}{2} \right)\right)\sin\left(\frac{y}{2} \right)\cos\left(\frac{y}{2} \right)}.
\end{equation}
There exists a generalization of the potential
$V_{\III}$, i.e.:
\begin{equation}
    V_{\III}^{\text{gen}}(r,\varphi) = \frac{\alpha}{r} + \frac{f(\varphi)}{r^{2}},
    \label{eqn:viiigen}
\end{equation}
where $f$ is an arbitrary function of $\varphi$.
Then the corresponding  equations are:
\begin{eqnarray}
        \ddot{r}&=&r\dot{\varphi}^2
        +\frac{\alpha}{r^{2}} + \frac{2 f(\varphi)}{r^{3}},
        \label{eqn:viiigeneqa}
        \\
       \ddot{\varphi}&=& - 2 \frac{\dot{r}}{r}\dot{\varphi}-
        \frac{f^\prime(\varphi)}{r^{2}},
        \label{eqn:viiigeneqb}
       \end{eqnarray}
where  prime indicates  the derivative of $f$ with respect to $\varphi$. Introducing the new variables $w_1,w_2,w_3,w_4$ as  in \eqref{eqn:syssubs} yields the following Hamilton equations:
\begin{eqnarray}
\dot w_1&=&w_3, \\[0.3cm]
\dot w_2 &=& \frac{w_4}{w_1^2},\\
\dot w_3 &=& \frac{\alpha w_1+2f(w_2)+w_4^2}{w_1^3},\\
\dot w_4 &=& -\frac{f'(w_2)}{w_1^2}.
\end{eqnarray}
We apply the reduction method \cite{kepler} by choosing $w_{2}=y$ as the new
independent variable, and consequently the following system of three
first-order equations is obtained:
\begin{eqnarray}
        \der{w_{1}}{y} &=& \frac{w_{3}w_1^2}{w_4},
        \label{eqn:v3geq_1}
         \\
        \der{w_{3}}{y}&=&\frac{\alpha w_1+2f(y)+w_4^2}{w_1w_4},
           \label{eqn:v3geq_2}     \\
       \der{w_{4}}{y} &=& -  \frac{f'(y)}{w_4}.
               \label{eqn:v3geq_3}
    \end{eqnarray}
Equation \eqref{eqn:v3geq_3} can be integrated to give
\begin{equation}
w_4=\pm \sqrt{J- 2f(y)},
\end{equation}
with $J$ an arbitrary constant.
Finally, eliminating $w_3$ by means of \eqref{eqn:v3geq_1}, i.e.
\begin{equation}
w_3=\frac{w_4}{w_1^2}\der{w_{1}}{y},
\end{equation}
 yields the following second-order equation for $w_1=w_1(y)$:
\begin{equation}
\frac{{\rm d}^2w_1}{{\rm d}y^2}=\displaystyle\frac{2 (J-2f(y))\left(\displaystyle\der{w_{1}}{y}\right)^2 + f'(y)w_1\displaystyle\der{w_{1}}{y} + \alpha w_1^3+ w_1^2J }
{w_1(J-2f(y))}.
    \label{eqn:u1eqgen}
\end{equation}
This equation  is  linearizable since it admits an eight-dimensional Lie symmetry algebra. As in the case of equation \eqref{eqn:w1fin},  the transformation \eqref{1su}
yields the linear equation:
\begin{equation}
\ddot u=\frac{f'(y)\dot u  - J u - \alpha}{J-2f(y)}.
\end{equation}

\section{The Tremblay-Turbiner-Winternitz system}
We now consider the superintegrable Tremblay-Turbiner-Winternitz (TTW) system \cite{TTW}, namely
an Hamiltonian system with a potential that generalizes $V_{\I}$ in \eqref{eqn:vi}, i.e.:
\begin{equation}
    V_{\text{TTW}}(r,\varphi) = \omega^{2}r^{2}
    + \frac{k^{2}}{r^{2}}\left( \frac{\beta_{1}}{\cos^{2}(k\varphi)}
    + \frac{\beta_{2}}{\sin^{2}(k\varphi)}\right).
    \label{eqn:ttw}
\end{equation}
The corresponding Lagrangian, i.e.:
\begin{equation}
L_{TTW}= \frac{1}{4}\left(\dot r^2+r^2\dot \varphi^2\right)-\omega^{2}r^{2}
    - \frac{k^{2}}{r^{2}}\left( \frac{\beta_{1}}{\cos^{2}(k\varphi)}
    + \frac{\beta_{2}}{\sin^{2}(k\varphi)}\right),\label{lagrTTW}
\end{equation}
yields the following Lagrangian equations:
\begin{subequations}
    \begin{align}
       \ddot{r}&=-4\omega^{2} r +r\dot\varphi^2+ \frac{4k^{2}}{r^{3}}
        \left( \frac{\beta_{1}}{\cos^{2}\left( k \varphi \right)}+
        \frac{\beta_{2}}{\sin^{2}\left( k \varphi \right)}
        \right),
        \label{eqn:ttweqa}
        \\
        \ddot{\varphi}&= - \frac{2}{r} \dot{r}\dot{\varphi}-
        \frac{4k^{3}}{r^{4}}
        \left( \frac{\beta_{1}\sin\left( k \varphi \right)}{%
            \cos^{3}\left( k \varphi \right)}-
            \frac{\beta_{2}\cos\left( k \varphi \right)}{%
            \sin^{3}\left( k \varphi \right)}
        \right),
        \label{eqn:ttweqb}
    \end{align}
    \label{eqn:ttweq}
\end{subequations}
that admit a three-dimensional
Lie point symmetry algebra $\Sl(2,\R)$ spanned by:
\begin{equation}
    \begin{gathered}
        \Sigma_{1}=\partial_{t}, \quad
        \Sigma_{2}=\cos(4\omega t)\partial_{t}
        - 2\omega\sin(4\omega t)r\partial_{r},
        \\
        \Sigma_{3}=\sin(4\omega t)\partial_{t}
        +2 \omega\cos(4\omega t)r\partial_{r},
            \end{gathered}    \label{eqn:sl2r}
\end{equation}
which are also Noether symmetries of the Lagrangian \eqref{lagrTTW}.
The application of Noether's theorem  yields three first integrals, one being the Hamiltonian, i.e.:
\begin{equation}H_{TTW}=\frac{1}{4}\left(\dot r^2+r^2\dot \varphi^2\right)+\omega^{2}r^{2}
    + \frac{k^{2}}{r^{2}}\left( \frac{\beta_{1}}{\cos^{2}(k\varphi)}
    + \frac{\beta_{2}}{\sin^{2}(k\varphi)}\right).\label{hamTTW}\end{equation}
The other two first integrals depend on $t$, i.e.:
\begin{eqnarray}
K_{2_{TTW}}&=&\left[\frac{1}{4}\left(\dot r^2+r^2\dot \varphi^2\right)-\omega^{2}r^{2}
    + \frac{k^{2}}{r^{2}}\left( \frac{\beta_{1}}{\cos^{2}(k\varphi)}
    + \frac{\beta_{2}}{\sin^{2}(k\varphi)}\right)\right]\cos(4\omega t)\nonumber\\
    &&+\omega r\dot r\sin(4\omega t),\\
K_{3_{TTW}}&=&  \left[\frac{1}{4}\left(\dot r^2+r^2\dot \varphi^2\right)-\omega^{2}r^{2}
    + \frac{k^{2}}{r^{2}}\left( \frac{\beta_{1}}{\cos^{2}(k\varphi)}
    + \frac{\beta_{2}}{\sin^{2}(k\varphi)}\right)\right]\sin(4\omega t)\nonumber\\
    &&-\omega r\dot r\cos(4\omega t).
    \end{eqnarray}
\begin{remark} Another time-independent first integral can
be obtained by the following combination:
\begin{equation}
 H_{TTW}^2-K_{2_{TTW}}^{2} - K_{3_{TTW}}^{2} =r^4\dot \varphi^2
    +4 k^{2}\left( \frac{\beta_{1}}{\cos^{2}(k\varphi)}
    + \frac{\beta_{2}}{\sin^{2}(k\varphi)}\right).
\end{equation}
\end{remark}
The presence of the algebra $sl(2,\mathbb{R})$ suggests to eliminate the two
parameters $\beta_1$ and $\beta_2$ by raising the order.   We solve system
\eqref{eqn:ttweq}  with respect to $\beta_1$ and $\beta_2$,
and then we take the derivative with respect to $t$  which yields the
following two third-order equations:
\begin{equation}
       r \dddot{r}+{\dot r}(16\omega^2r + 3 \ddot{r})=0,
        \label{eqn:ttwhighera}
\end{equation}
\begin{eqnarray}
         \cos(k\varphi)\sin(k\varphi)r^2\dddot{\varphi}
        + 3\cos^{2}(k\varphi) k r^2 \dot{\varphi}\ddot{\varphi}
       + 6\cos^{2}(k\varphi)k r \dot{r}\dot{\varphi}^2\nonumber \\
      + 16\cos(k\varphi)\sin(k\varphi) k^2 \omega^2 r^2\dot{\varphi}
                 - 4\cos(k\varphi)\sin(k\varphi)k^2 r^2\dot{\varphi}^3
        + 4\cos(k\varphi)\sin(k\varphi) k^2 r\ddot{r}\dot{\varphi}\nonumber \\
       + 6\cos(k\varphi)\sin(k\varphi)r\dot{r}\ddot{\varphi}
              + 2\cos(k\varphi)\sin(k\varphi)r\ddot{r}\dot{\varphi} \nonumber \\
       + 6\cos(k\varphi)\sin(k\varphi)\dot{r}^2\dot{\varphi}
        - 3\sin^{2}(k\varphi) k r^2 \dot{\varphi}\ddot{\varphi}
       - 6\sin^{2}(k\varphi) k r\dot{r}\dot{\varphi}^2=0.
         \label{eqn:ttwhigherb}
 \end{eqnarray}
Equation  \eqref{eqn:ttwhighera} admits a seven-dimensional Lie symmetry algebra generated
by the following operators:
\begin{eqnarray} X_1=\partial_t,\; X_2=\cos(4\omega t)\partial_t-2\omega\sin(4\omega
t)r\partial_r,\;
 X_3=\sin(4\omega t)\partial_t+2\omega\cos(4\omega t)r\partial_r,\nonumber\\
 X_4=\frac{\cos(4\omega t)}{r}\partial_r,\;
 X_5=\frac{\sin(4\omega t)}{r}\partial_r,\;
 X_6=r\partial_r,\;X_7=\frac{1}{r}\partial_r,\end{eqnarray}
and consequently it is linearizable. We find that a two-dimensional non-abelian intransitive subalgebra is that generated by $X_6$ and $X_7$, and following Lie's classification \cite{Lie12}, if we transform them into their  canonical form, i.e., $\partial_u, u\partial_u$, then  we obtain that  the new dependent variable is given by $$u=\frac{r^2}{2}$$ and consequently equation  \eqref{eqn:ttwhighera} becomes \begin{equation}\dddot u=-16\omega^2\dot u,\end{equation} namely  the derivative with respect to $t$ of the equation of  a linear harmonic oscillator with frequency $4\omega$. Thus, the general solution of \eqref{eqn:ttwhighera} is
\begin{equation} r=\sqrt{a_1+a_2\cos(4\omega t)+a_3\sin(4\omega t)}.\label{rsol}\end{equation}
 Equation \eqref{eqn:ttwhigherb} is also linearizable since it admits a seven-dimensional Lie symmetry algebra generated by
\begin{equation} \Omega = s_1(t)
\partial_{t} + \frac{ - \cos^{2}(k\varphi) s_2(t) + 2 k s_3(t)}{2 \cos(k
\varphi) \sin(k \varphi)k}
\partial_{\varphi},\end{equation}
with $s_1, s_2, s_3$ that satisfy the following seventh-order linear system:
\begin{subequations}
    \begin{align}
        &r^2\dddot{s}_{1}
        +\begin{aligned}[t]
            &4 \dot{s}_{1} \ddot{r} k^2 r - 4\dot{s}_{1} \ddot{r} r
            + 16 \dot{s}_{1} k^2 \omega^2 r^2 - 8 \ddot{r} \dot{r} k^2 s_{1}
            \\
            &+ 8 \ddot{r} \dot{r} s_{1}
            - 32 \dot{r} k^2 \omega^2 s_{1} r + 32 \dot{r} \omega^2 s_{1} r=0,
        \end{aligned}
        \\
        &r^2\dot{s}_{2}-\ddot{s}_{1} r^{2} + 2 \dot{s}_{1} r \dot{r}
        + 2 r\ddot{r} s_{1}  -2 \dot{r}^2 s_{1} = 0,
        \\
        &r^{2}\dddot{s}_{3}+6 \ddot{s}_{3} \dot{r} r + 4 \dot{s}_{3} \ddot{r} k^2 r
        + \dot{s}_{3} \ddot{r} r + 6 \dot{s}_{3} \dot{r}^2 + 16 \dot{s}_{3} k^2 \omega^2 r^2=0,
    \end{align}
    \label{eqn:deteqttwdphi}
\end{subequations}
with $r$ given  in \eqref{rsol}.
Similarly to  equation  \eqref{eqn:ttwhighera}, we find that a  two-dimensional non-abelian intransitive subalgebra is generated by the operators \begin{equation}-\frac{1}{2k}\cot(k\varphi)\partial_{\varphi}, \quad \frac{2}{\sin(2k\varphi)}\partial_{\varphi},\end{equation} that put into canonical form yield the new dependent variable $$ v=-\frac{1}{2k}\cos^2(k\varphi),$$ and consequently
 equation \eqref{eqn:ttwhigherb} becomes linear, i.e.:
\begin{equation} \dddot v=-\frac{6\dot r}{r} \ddot v-\frac{2}{r^2}\left(3\dot r^2 + 8k^2\omega^2r^2+\left(2k^2 + 1\right)r\ddot r\right)\dot v.\label{TTWveq}\end{equation}
\begin{remark} The TTW system admits closed orbits if $k$ is rational, as it has been shown by various methods in \cite{TTW10}, \cite{KKM10}, \cite{GoneraPLA12}. We observe that equation \eqref{TTWveq} yields solutions of \eqref{eqn:ttweq} in terms of hypergeometric and trigonometric functions if $k$ is rational, although the linearization that we have achieved
 remains valid even for $k$ irrational.  \end{remark}

\section{Conclusions}
In this paper we have considered  superintegrable systems  in  two-dimensional real
Euclidean space, and shown that they possess hidden symmetries leading to linearization.

Other superintegrable systems have been found in
two-dimensional non-Euclidean spaces, i.e. in two-dimensional space with
non-constant curvature. Examples of such systems are the Perlick system
\cite{Perlick}, the Taub-NUT system \cite{Manton},  superintegrable systems for
the Darboux space of Type I \cite{KalKreWint}, and others \cite{Riglioni11}, \cite{Latini15}.

In a forthcoming paper \cite{2Dnonflat} we will show  that also
superintegrable systems in non-Euclidean space can be reduced to linear
equations by means of their hidden symmetries.

\section*{Acknowledgement}
MCN acknowledges the support of the Italian Ministry of University and
Scientific Research through PRIN 2010-2011, Prot. 2010JJ4KPA\_004, ``Geometric
and analytic theory of Hamiltonian systems in finite and infinite dimensions".

\end{document}